\documentclass[prl,
twocolumn,
bibnotes,
 amsmath,amssymb,
 aps,
]{revtex4-2}

\usepackage{graphicx}
\usepackage{dcolumn}
\usepackage{bm}
\usepackage[utf8]{inputenc}
\usepackage{todonotes}
\usepackage{comment}
\usepackage[colorlinks,linkcolor=red,citecolor=blue,urlcolor=blue]{hyperref}

\usepackage{xcolor}

\usepackage{amsmath}
\usepackage{amssymb}
\usepackage{bm}

\usepackage{soul}

\newcommand{\ikr}[1]{{\color{red}{#1}}}

\begin{document}

\title{Practical Kinetic Models for Dense Fluids}

\author{Ilya Karlin}
\email{ikarlin@ethz.ch}\thanks{Corresponding Author}
\affiliation{Department of Mechanical and Process Engineering, ETH Z{u}rich, 8092 Z{u}rich, Switzerland}

\author{Seyed Ali Hosseini}
\email{shosseini@ethz.ch}
\affiliation{Department of Mechanical and Process Engineering, ETH Z{u}rich, 8092 Z{u}rich, Switzerland}

\date{\today}

\begin{abstract}
   Nonlinear idempotent operator instead of a linear projection is introduced to derive kinetic models for dense fluids. A new lattice Boltzmann model for compressible two-phase flow is derived based on the Enskog--Vlasov kinetic equation as an example of practical importance. 
\end{abstract}
\maketitle

Starting with pioneering works by Mori \cite{mori_1958_PhysRev.112.1829} and Zwanzig \cite{zwanzig_1961_PhysRev.124.983}, projection operator techniques constitute a basis for modern statistical mechanics and kinetic theory \cite{grabert_1982,BALIAN19861}. 
Our concern here is the application of projector operator techniques to produce kinetic models of practical utility for dense fluids simulations. To that end, perhaps one of the the earliest such application was the work of Dufty, Santos and Bray \cite{dufty_1996_PhysRevLett.77.1270} in the context of a kinetic model for the Enskog transport equation for hard spheres. The recent surge of interest for this kind of kinetic models can be seen in a number of works, e.\ g. \cite{huang2021mesoscopic,shan_2025,busuioc_2025,su_2023_PhysRevFluids.8.013401,Shan_Su_Gibelli_Zhang_2023}.
Here we  show that changing the linear projection paradigm to a more general {nonlinear idempotent} operator provides a way to kinetic models of practical utility. As an example, we derive a new explicit lattice Boltzmann model for compressible two-phase flow, based on the Enskog--Vlasov kinetic equation and by using gauge invariance of the hydrodynamic limit with respect to the thermodynamic pressure. 

Starting point is the first BBGKY equation,
\begin{equation}
    \label{eq:BBGKY}
    \begin{split}
 &   \partial_t f+\bm{v}\cdot\bm{\nabla}f=\mathcal{J},\\
 &   J=\int\int \bm{\nabla} V\left(\lvert \bm{r}-\bm{r}'\rvert\right)\cdot\frac{\partial}{\partial\bm{v}}f_2(\bm{r},\bm{v},\bm{r}',\bm{v}',t)d\bm{v}'d\bm{r}',
 \end{split}
\end{equation}
 where $f(\bm{r},\bm{v},t)$ and $f_{2}(\bm{r},\bm{v},\bm{r}',\bm{v}',t)$ are the one- and the two-particle distribution functions, respectively, $\bm{r}$, $\bm{r}'$ and $\bm{v}$, $\bm{v}'$ are particles position and velocity, while $V$ is a pair interaction potential. Note that the BBGKY equation satisfies the local mass conservation, 
   $  \int\mathcal{J}d\bm{v}=0$.
The local equilibrium state is defined by the Maxwellian $f^{\rm eq}$, parameterized by the local values of number density $n$ and flow velocity $\bm{u}$ and temperature $T$, 
    \begin{equation}\label{eq:LM}
    	f^{\rm eq}(\bm{u},RT)=\frac{n}{\left(2\pi RT\right)^{3/2}}\exp\left[-\frac{(\bm{v}-\bm{u})^2}{2RT}\right],
    \end{equation}
where $R$ is gas constant.
Let us define the non-local force and the rate of the kinetic energy change,
\begin{align}
    &\bm{F}[J]=\int m\bm{v}\mathcal{J}d\bm{v},\label{eq:defF}\\
     &{\tilde{Q}[J]=\int \frac{m(\bm{v}-\bm{u})^2}{2}\mathcal{J}d\bm{v},}\label{eq:defQ}
\end{align}
and consider the following operation applied on the right hand side of the BBGKY equation,
\begin{equation}\label{eq:K}
  \mathcal{K}_{\lambda}(J)=\frac{1}{\lambda}\left[f^{\rm eq}\left(\bm{u}_{\lambda}^*, RT^*_{\lambda}\right)-f^{\rm eq}(\bm{u},RT)\right],
\end{equation}
with the shifted flow velocity $\bm{u}^*_{\lambda}$ and the shifted temperature $T^*_{\lambda}$ defined as,
\begin{align}
    \label{eq:def_ustar}
&    \bm{u}_{\lambda}^*=\bm{u}+\frac{\lambda}{\rho}\bm{F},\\
& {RT^*_{\lambda}=RT+\frac{2\lambda}{3\rho}\tilde{Q}
-\frac{\lambda^2}{3\rho^2}\bm{F}\cdot\bm{F}.}
\end{align}
Here $m$ is the mass of the particle, $\rho=mn$ is the fluid density, and $\lambda>0$ is a parameter with the dimension of the time which will be specified below.
Note that, even though the operator \eqref{eq:K} is not a linear projection, it nevertheless features the idempotency property,
    $\mathcal{K}_{\lambda}(\mathcal{K}_{\lambda}(\mathcal{J}))=\mathcal{K}_{\lambda}(\mathcal{J})$.
Moreover, operation \eqref{eq:K} satisfies the {orthogonality} condition,
$    \mathcal{K}_{\lambda}(\mathcal{J}-\mathcal{K}_{\lambda}(\mathcal{J}))=0$.
Thus, much like the conventional linear projection operator, the nonlinear idempotent operation \eqref{eq:K} can be applied to split the  two-particle interaction $J$ into the local and non-local contributions,
    $\mathcal{J}=\mathcal{J}_{\rm loc}+\mathcal{J}_{\rm nloc}$, where
    $\mathcal{J}_{\rm loc}=\mathcal{J}-\mathcal{K}_{\lambda}(\mathcal{J})$ and
    $\mathcal{J}_{\rm nloc}=\mathcal{K}_{\lambda}(\mathcal{J})$.
The part $\mathcal{J_{\rm loc}}$ satisfies the local conservation of mass, momentum and kinetic energy,
  $  \int \{1,\bm{v}, v^2\}\mathcal{J}_{\rm loc}d\bm{v}=0$.
Hence, the simplest rationale for the local part is a single--relaxation--time Bhatnagar--Gross--Krook (BGK) approximation, also used previously in a similar context \cite{dufty_1996_PhysRevLett.77.1270}, 
    $\mathcal{J}_{\rm loc}\to -({1}/{\tau})(f-f^{\rm eq})$,
where $\tau$ is the \emph{bare} relaxation time to the local equilibrium. 
In order to save notation, we write,
and refer to the distribution function $f^*=f^{\rm eq}(\bm{u}^*_\lambda,RT^*_\lambda)$ in \eqref{eq:K} as shifted equilibrium.
Summarizing, the proposed kinetic model takes the form,
\begin{equation}
    \label{eq:final_kinetic_model}
    \partial_t f+\bm{v}\cdot\bm{\nabla}f=-\frac{1}{\tau}(f-f^{\rm eq})+\frac{1}{\lambda}(f^*-f^{\rm eq}).
\end{equation}
Comments are in order:\\
\noindent (i) In order to gain some intuition about relative magnitude of the relaxation parameters in \eqref{eq:final_kinetic_model}, it is instructive to rewrite the right hand side of Eq.\ \eqref{eq:final_kinetic_model} as follows:
\begin{equation}\label{eq:final_kinetic_model2}
	\partial_t f + \bm{v}\cdot\bm{\nabla} f = -\frac{1}{\theta}\left(f-f^{\rm eq}\right) 
    - \frac{1}{\lambda}(f-f^{*}),
\end{equation}
where we have introduced the \emph{effective} relaxation time to the equilibrium, 
    $\theta={\tau\lambda}/({\lambda-\tau})$.
The form \eqref{eq:final_kinetic_model2} reveals the two relaxation processes executed in parallel: Relaxation to the equilibrium with the effective relaxation time $\theta$ and the relaxation to the shifted-equilibrium with the relaxation time $\lambda$. Positivity of the effective relaxation time $\theta\ge 0$ requires a hierarchy of relaxation times to hold, 
 $   \label{eq:hierarchy}\lambda\ge \tau$.
Moreover, the effective relaxation time $\theta$ converges to the bare relaxation time $\tau$ when the relaxation to the equilibrium becomes dominant,
$    \theta\to\tau\ {\rm for}\ \lambda\gg\tau$.
\\
\noindent (ii) In the limit $\lambda\to0$, the nonlinear idempotent operator \eqref{eq:K} becomes a familiar linear projection \cite{dufty_1996_PhysRevLett.77.1270,hosseini2022towards},
\begin{equation}
    \label{eq:lim_projection}
    \begin{split}
   \lim_{\lambda\to 0}\mathcal{K}_{\lambda}(J)=\frac{1}{\rho}\frac{\partial f^{\rm eq}}{\partial\bm{u}}
\cdot \bm{F}[J]
   +\frac{2}{3\rho}\frac{\partial f^{\rm eq}}{\partial (RT)}\tilde{Q}[J].
   %
   \end{split}
\end{equation}
{In other words, the essence of the proposed formalism is to {implement the nonlocal effects in the form of a relaxation} \eqref{eq:final_kinetic_model2} towards the conventional force term \eqref{eq:lim_projection}, with a "slow" relaxation time $\lambda$. Therefore, it is expected (and will be confirmed in the following) that the transport coefficients in the hydrodynamic limit shall depend on the bare relaxation time $\tau$ rather than on $\lambda$.}\\
\noindent (iii) The proposed kinetic model has a special advantage when it comes to space-time discretization for a numerical realization. To that end, the method of integration over characteristics as introduced in \cite{he_1998_novel} results in the following second-order accurate in time scheme (see details in the Supplemental Materials):
\begin{equation}
\begin{split}
    \label{eq:pre_LBM}
    f(\bm{r}+\bm{v}\delta t,t+\delta t)={f}+ 2\beta\left(f^{\rm eq} - {f}\right)+ {\frac{\delta t}{\lambda}}\left(1-\beta\right) \left(f^* - f^{\rm eq}\right),
    \end{split}
\end{equation}
where the right-hand side is evaluated at time $t$ and position $\bm{r}$, where $\delta t$ is the time step, and
where $\beta\in[0,1]$ is tight to the bare relaxation time, 
\begin{equation}\label{eq:beta}
\beta=\frac{\delta t}{2\tau+\delta t}.
\end{equation}
Moreover, the flow velocity $\bm{u}$ and the temperature $T$ entering the equilibrium and the shifted equilibrium distributions elsewhere in \eqref{eq:pre_LBM} are computed as,
\begin{align}
&\rho=\int mfd\bm{v},\\
\label{eq:u_shift}
  &  \rho\bm{u}=\int m\bm{v}fd\bm{v} + \frac{\delta t}{2}\bm{F},\\
  &  {\frac{3}{2}\rho RT= \int \frac{m(\bm{v}-\bm{u})^2}{2}fd\bm{v} + \frac{\delta t}{2}\tilde{Q}.}\label{eq:T_shift}
\end{align}
Finally, we set $\lambda=\delta t$, which is consistent with the requirement $\tau\ll\lambda$ for $\beta\to1$.




Evaluation of the non-local force \eqref{eq:defF} and of the  kinetic energy rate \eqref{eq:defQ} requires to specify the particle interaction. 
It is customary to invoke the Enskog--Vlasov model \citep{enskog_warmeleitung_1921,vlasov_many-particle_1961,grmela_approach_1974,karkheck_kinetic_1981} where the hard-sphere collisions and a weak long-range attraction potential contribute to a non-local momentum and energy transfer. 
    Evaluation 
    proceeds along familiar lines \citep{chapman_mathematical_1939,he_thermodynamic_2002,he1998discrete,martys2006bbgky} and results in the following lower-order approximation, sufficient to our purpose of addressing the hydrodynamic limit; see Supplemental Materials: 
    \begin{align}
    &{\bm{F}= -
	\bm{\nabla}\left(P_{\rm E}-P_0\right)
	-  \rho\bm{\nabla}V_{\rm V},}
	\label{eq:EVforce1}\\
    &\tilde{Q}=-(P_{\rm E}-P_0)(\bm{\nabla}\cdot\bm{u}).\label{eq:Ework}
    \end{align}
    {Here $P_{\rm E}$ is the Enskog's pressure which we assume in a van der Waals form, $P_{\rm E}=\rho RT/(1-b\rho)$, with $b$ the excluded volume parameter,}
    and $P_0=\rho RT$ is the ideal gas pressure. Furthermore,
    $V_{\rm V}=\int V\left(\lvert \bm{r}-\bm{r}'\rvert\right)\rho(\bm{r}',t) d\bm{r}'$ is Vlasov's mean field potential;
     The corresponding Vlasov's force shall be approximated as $\bm{F}_{\rm V}=-\rho\bm{\nabla}V_{\rm V}=-\bm{\nabla}P_{\rm V}+\kappa\rho\bm{\nabla}\bm{\nabla}\cdot\bm{\nabla}\rho$, where $P_{\rm V}=-a\rho^2$ is the pressure contribution of the long-range attractive force while $\kappa$ is the capillarity coefficient. 
     The rate of kinetic energy change due to collisional Enskog's transport manifests in the corresponding work of compression \eqref{eq:Ework}. 
While the above low-order approximations are standard, 
we emphasize a strong coupling between the momentum and energy terms in the transformations \eqref{eq:u_shift} and \eqref{eq:T_shift}. Indeed, unlike the familiar isothermal case where the force \eqref{eq:EVforce1} is solely a function of the density, in the compressible case the coupling is also due to the non-uniformity of the temperature and the non-uniformity of the flow in the compression work \eqref{eq:Ework} which makes the transformation \eqref{eq:u_shift} and \eqref{eq:T_shift} implicit. We shall return to this issue below.

\begin{figure}[t!]
	    \centering
		    \includegraphics[width=0.8\columnwidth]{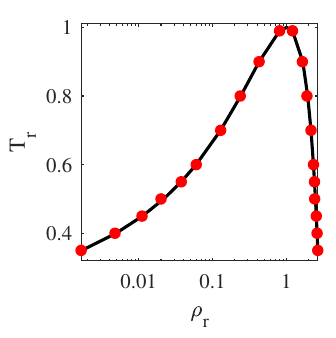}
	    \caption{Liquid-vapor coexistence for the van der Waals equation of state, with the reference pressure $P_0=P_E$. Axes are reduced by critical values of density and temperature. Line: Maxwell's equal-area construction; Symbol:  Simulation.
	    }
	    \label{Fig:Coexistence}
    \end{figure}

Discrete equations of the form \eqref{eq:pre_LBM} become particularly useful for numerical simulation when the velocity is also rendered  discrete. This is the case of the lattice Boltzmann realization considered below
%
%
%
%
using the standard discrete velocity set $D3Q27$, where $D=3$ stands for three dimensions and $Q=27$ is the number of discrete velocities, 
    	$\bm{c}_i=(c_{ix},c_{iy},c_{iz})$, $c_{i\alpha}\in\{-1,0,1\}$, and we merely need to specify the corresponding equilibrium and shifted equilibrium.
{We first address the mass and momentum transport.}
Let us define functions in two variables, $\xi_{\alpha}$ and $\zeta_{\alpha\alpha}$, 
    \begin{equation}
        \label{eq:phi}
 \Psi_{{i\alpha}}(\xi_{\alpha},\zeta_{\alpha\alpha})=1-c_{i\alpha}^2+ \frac{1}{2}\left[(3c_{i\alpha}^2-2)\zeta_{\alpha\alpha}+c_{i\alpha}\xi_{\alpha}\right].
    \end{equation}
All pertinent populations shall be constructed using products of functions \eqref{eq:phi}. Specifically,  the equilibrium populations $f_i^{\rm eq}$ are defined by setting $\xi_{\alpha}^{\rm eq}=u_\alpha$ and $\zeta_{\alpha\alpha}^{\rm eq}=P_0/\rho+u_{\alpha}^2$,
\begin{equation}
    \label{eq:feq}
    f_i^{\rm eq}=\rho\prod_{\alpha}\Psi_{{i\alpha}}\left(u_\alpha,\frac{P_0}{\rho}+u_{\alpha}^2\right).
\end{equation}
For the shifted equilibrium, we set: $\xi_\alpha^*=u_\alpha+(\lambda/\rho)F_{\alpha}$ and $\zeta_{\alpha\alpha}^{*}=P_0/\rho+{(u_{\alpha} + (\lambda/\rho)F_\alpha)}^2 +(\lambda/\rho){\Phi_{\alpha\alpha}}$ to get,
\begin{equation}
    \label{eq:f_star}
    f_i^{*}=\rho\prod_{\alpha}\Psi_{{i\alpha}}\left(u_\alpha+\frac{\lambda}{\rho}F_{\alpha},\zeta_{\alpha\alpha}^{*}\right),
\end{equation}
with the non-equilibrium momentum flux correction as,
\begin{equation}\label{eq:correction_u}
    \Phi_{\alpha\alpha}=\partial_{\alpha}\left(\rho u_{\alpha} \left(u_{\alpha}^2 + \frac{3P_{0}}{\rho}-3\varsigma^2\right)\right)
    + \Phi'.
\end{equation}
Here $\varsigma=1/\sqrt{3}$ is the lattice speed of sound, while the second term reads,
\begin{equation}\label{eq:correction_gamma}
    \Phi' = P_0\left(\frac{5}{3} - \gamma_0\right)
    (\bm{\nabla}\cdot\bm{u}),
\end{equation}
where $\gamma_0$ is a pseudo-adiabatic parameter,
\begin{equation}\label{eq:gamma}
    {\gamma_0=\left(\dfrac{\partial \ln P_0}{\partial \ln\rho}\right)_T+\frac{2}{3}\left(\frac{P_{\rm E}}{\rho R T}\right)\left(\dfrac{\partial \ln P_0}{\partial \ln T}\right)_\rho}.
\end{equation}
 The resulting lattice Boltzmann equation follows the pattern of Eq.\ \eqref{eq:pre_LBM} upon replacement $f\to f_i$, whereby the left hand side becomes exact on-lattice propagation while the flow velocity elsewhere in the equilibrium and shifted equilibrium populations \eqref{eq:feq} and \eqref{eq:f_star} is computed as in \eqref{eq:u_shift}, with summation over discrete velocities replacing integration. Several comments are in order:\\
(i) The proposed lattice Boltzmann equation replicates the general kinetic model \eqref{eq:pre_LBM} while adopting the necessary adjustments due to the minimalist discrete velocity setting.
First term in \eqref{eq:correction_u} is a correction of viscous stress due a bias of the discrete velocities, $c_{i\alpha}^3=c_{i\alpha}$. The second correction term \eqref{eq:correction_gamma} maintains the vanishing bulk viscosity of the Enskog--Vlasov fluid.
With this, the continuity and the momentum equations are recovered as follows (see Supplementary Materials for details),
\begin{align}\label{eq:macro_mass}
      &  \partial_t\rho + \bm{\nabla}\cdot\rho \bm{u} = 0,\\
       &  {\partial_t\rho\bm{u} + \bm{\nabla}\cdot\rho \bm{u}\otimes\bm{u} 
       +\bm{\nabla}P_{\rm E}+\rho \bm{\nabla}V_{\rm V}
       + \bm{\nabla}\cdot\bm{T}_{\rm NS}= 0,}\label{eq:momentum_balance}
    \end{align}
   where $\bm{T}_{\rm NS}=-\mu\bm{S}$ is
the Navier--Stokes viscous stress tensor
with $\bm{S}=\bm{\nabla}\bm{u} + {\bm{\nabla}\bm{u}}^{\dagger} -({2}/{3})(\bm{\nabla}\cdot\bm{u})\bm{I}$ the trace-free rate-of-strain tensor,
while $\mu$ is the viscosity, 
\begin{align}
    &	\mu= \left(\frac{1}{2\beta} - \frac{1}{2}\right)\delta t P_0. \label{eq:viscosity}
    \end{align}
(ii) Note that, as anticipated, up to second order, the hydrodynamic limit does not depend on the relaxation parameter $\lambda$, while the sole dependence on the "fast" relaxation time $\tau$ is manifest through the dependence on the parameter $\beta$ \eqref{eq:beta} in the transport coefficient \eqref{eq:viscosity}.\\ 
(iii) Finally, and most importantly, the recovered momentum equation \eqref{eq:momentum_balance} is form-invariant with respect to the reference pressure $P_0$. This allows to consider the latter as a parameter that can be conveniently chosen, and not necessarily set to be the ideal gas pressure. As an interim example, Fig.\ \ref{Fig:Coexistence} shows the liquid-vapor coexistence diagram for the case $P_0=P_{\rm E}$ (this choice of reference pressure shall be explained shortly).


Moving onto the energy dynamics,
 we adopt a double-population path, following the method of \cite{karlin_consistent_2013,saadat_extended_2021,sawant_consistent_2021}.
%
%
%
To that end, the \emph{total kinetic energy}, $\rho E=\frac{3}{2}\rho R T+\frac{1}{2}\rho u^2$ is used as the locally conserved moment of the populations $g_i$, and 
operators $\mathcal{O}_\alpha$ acting on a smooth function $A(\bm{u})$ are introduced as,
\begin{equation}
    \mathcal{O}_{\alpha} A = \left(\frac{P_0}{\rho}\right)
    \frac{\partial A}{\partial  u_\alpha} + u_\alpha A.
\end{equation}
By setting symbols $\xi_{\alpha}=\mathcal{O}_{\alpha}$, $\zeta_{\alpha\alpha}=\mathcal{O}^2_{\alpha}$ in functions \eqref{eq:phi}, equilibrium populations $g^{\rm eq}_i$ are defined as the application of product-operators on the generating function $E(\bm{u},RT)$,
\begin{align}
&g_i^{\rm eq}(\bm{u},RT)=\rho\prod_{\alpha}\Psi_{{i\alpha}}\left(\mathcal{O}_\alpha,\mathcal{O}_{\alpha}^2\right)E({\bm{u},RT}).\label{eq:geq}
\end{align}
Shifted equilibrium populations are defined as,
\begin{align}
&g_i^{*}=g_i^{\rm eq}(\bm{u}^*_{\lambda},RT^*_{\lambda})+g_i',
\label{eq:gstar}
\end{align}
with the correction term $g_i'$,
\begin{align}
		g_{i}'= \left\{\begin{aligned}
			& ({1}/{2})\bm{c}_i\cdot\bm{q}', &\text{ if } c_i^2=1, & \\ 
			&0, & \text{otherwise}.&\\
		\end{aligned}\right.
	\label{eq:g_corr}	
	\end{align}
Here the non-equilibrium energy flux correction is the vector $\bm{q}'$ with the components,
\begin{equation}\label{eq:correction_q}
    {q}_{\alpha}' = \lambda P_0\partial_{\alpha}\left(\frac{P_0}{\rho} - RT\right) + \lambda {u}_{\alpha}{\Phi}'.
\end{equation}
First term in \eqref{eq:correction_q} provides a correction to the Fourier heat flux while the second term is required for consistency with the above bulk viscosity correction \eqref{eq:correction_gamma}. The lattice Boltzmann equation for the $g_i$ populations follows again the pattern of \eqref{eq:pre_LBM}, with a replacement $f\to g_i$ and the total kinetic energy balance equation is recovered as follows (see details in the Supplemental Materials):
 
\begin{multline}
    \partial_t\rho E + \bm{\nabla}\cdot \rho \bm{u}E + \bm{u}\cdot\left(\bm{\nabla}P_{\rm E}+\rho\bm{\nabla}V_{\rm V}\right) +P_{\rm E}(\bm{\nabla}\cdot\bm{u})\\ + \bm{\nabla}\cdot\left(\bm{u}\cdot\bm{T}_{NS}\right) - \bm{\nabla}\cdot(k\bm{\nabla}T) =0. \label{eq:LB_energy_balance}
\end{multline}
%
where $k$ is the thermal conductivity, 
 $k= ({5}/{2})R\mu$.
{Comments are in order:\\
(i) 
Also the total kinetic energy balance equation \eqref{eq:LB_energy_balance} is {form-invariant} with respect to the reference pressure $P_0$, same as we observed it earlier with the momentum equation \eqref{eq:momentum_balance}. This allows to set the \emph{gauge} $P_0$ based on convenience.\\
(ii) At a first glance, it seems natural to choose the ideal gas gauge, $P_0=\rho RT$, as this annihilates the Fourier flux correction in \eqref{eq:correction_q}. However, a closer look reveals that the transformation relations \eqref{eq:u_shift} and \eqref{eq:T_shift} remain strongly coupled and result in an \emph{implicit} problem of elliptic type. This makes the lattice Boltzmann scheme impractical as it deprives it from the explicit realization, and we note again that this is in a marked contrast to the familiar isothermal case.\\
(iii) The gauge that renders the transformations \eqref{eq:u_shift} and \eqref{eq:T_shift} \emph{explicit} is the Enskog's gauge,
\begin{equation}
    \label{eq:P0E}
    P_0=P_{\rm E}.
\end{equation}
Indeed, with $P_0=P_{\rm E}$, the kinetic energy rate cancels in \eqref{eq:Ework} while the temperature dependence cancels in the non-local force \eqref{eq:EVforce1}, which renders the system \eqref{eq:u_shift} and \eqref{eq:T_shift} solvable. 
}

\begin{figure}[t!]
	    \centering
		    \includegraphics[width=0.8\columnwidth]{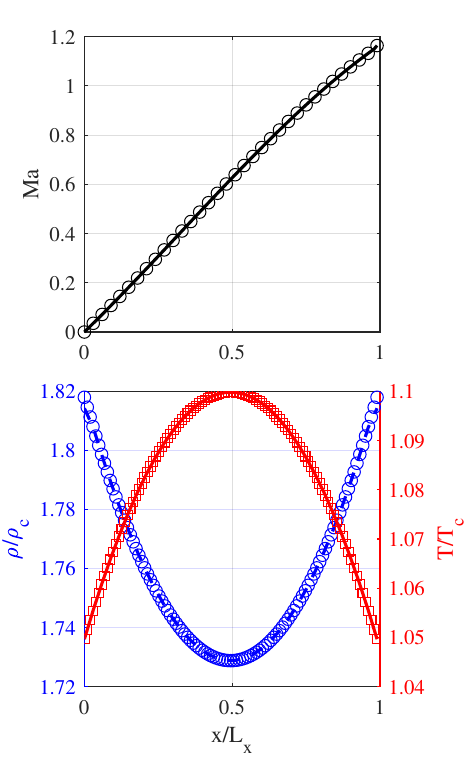}
	    \caption{ Top panel: Distribution of Mach number along the channel for the thermal Couette flow. Reference pressure is set to $P_0=P_E$. line: simulations, markers: analytical solution. Bottom panel: Distribution of density and temperature along the channel. Red filled and blue dashed lines are numerical results for temperature and density. Red square marker and blue circular marker are corresponding analytical solutions.
	    }
	    \label{Fig:couette_vdW_PRL}
    \end{figure}

With the choice of the gauge \eqref{eq:P0E}, the fully explicit lattice Boltzmann model of the Enskog--Vlasov fluid is complete. For a numerical validation, we consider a super-sonic thermal Couette flow in a two-dimensional channel of width $L_x$ between  a static and a moving wall, defined via the following boundary conditions,
	$\{u_x, u_y, T\}= \{0, 0, T_w\}$ at $x=0$ and 
    $\{u_x, u_y, T\} = \{0, U_w, T_w\}$ at $x=L_x$,
where $T_w$ is the fixed temperature of both walls and $U_w$ the speed of the moving wall. At steady state, once energy loss at the walls is balanced by energy production due to viscous heating a steady state solution is reached.
A simulation with the compressible model was run for ${\rm Ma}=1.2$ at the moving wall. Results are compared to the analytical solution in Fig. \ref{Fig:couette_vdW_PRL} and point to excellent agreement.

{In summary, we proposed a novel strategy to derive kinetic models of a practical utility from microscopic statistical theories. This is achieved by applying a new nonlinear idempotent operator instead of conventional linear projection to split the local and nonlocal contributions of the particles' interaction. The universal construction was specified for the classical Enskog--Vlasov theory to derive a particularly important lattice Boltzmann model for compressible non-ideal fluid. The derivation, targeting the hydrodynamic limit, revealed a gauge-invariance of the theory with respect to the choice of a reference pressure. While a similar invariance is merely a matter of convenience in the familiar isothermal situation \cite{hosseini2022towards}, we have shown that in the non-ideal compressible case the requirement of explicitness (and thus causality) of the lattice Boltzmann scheme is a much more stringent constraint, which requires selecting Enskog's excluded volume contribution to the thermodynamic pressure as the reference pressure. The resulting lattice Boltzmann system on the standard minimalist lattice is almost as simple and efficient as the classical. Due to its simplicity, there are many directions the present model can be further developed. Since we here used the simplest BGK local relaxation, the usual restriction on the Prandtl number can be removed by applying a Shakhov \cite{shakhov1968generalization} or quasi-equilibrium \cite{ansumali_2007_quasi} kinetic models. The proposed universal framework of implementing complex flow physics in the lattice Boltzmann setting can be extended to multi-component fluids \cite{sawant_consistent_2021}. Overall, the present new approach finalizes the development of kinetic models of practical utility beyond the standard lattice Boltzmann model.}
\section*{Acknowledgement}
This work was supported by European Research Council (ERC) Advanced Grant No. 834763-PonD and by the Swiss National Science Foundation (SNSF) Grants 200021-228065 and 200021-236715. Computational resources at the Swiss National Super Computing Center (CSCS) were provided under Grants No. s1286 and sm101. 
\bibliographystyle{apsrev4-1}
\bibliography{references}
\end{document}